\pdfoutput=1
\documentclass[
 aps,
 pra,
 reprint,
 amsmath,
 amssymb,
 superscriptaddress,
 floatfix,
 nofootinbib
]{revtex4-2}

\usepackage{graphicx}
\usepackage{bm}
\usepackage{color}
\usepackage[breaklinks=true,colorlinks=true,linkcolor=blue,citecolor=blue,urlcolor=blue]{hyperref}

\graphicspath{{figures/}}

\newcommand{\ud}{\mathrm{d}}

\newcommand{\uv}{\bm{u}}
\newcommand{\CEP}{\varphi_{0}}

\begin{document}

\title{Rapidity-Coupled Spin Dynamics in Pulsed Laser Fields \\
from Physics-Informed Neural Networks}

\author{N.~S. Akintsov}
\email[Corresponding author: ]{akintsov777@ntu.edu.cn}
\altaffiliation[ORCID: ]{0000-0002-1040-1292}
\affiliation{School of Artificial Intelligence and Computer Science,
Nantong University, Nantong 226019, China}

\author{A.~P. Nevecheria}
\email{artiom.nevecherya@gmail.com}
\altaffiliation[ORCID: ]{0000-0001-6736-4691}
\affiliation{Department of Mathematical and Computer Methods,
Kuban State University, Krasnodar 350040, Russia}

\author{S.~N. Andreev}
\email{andreev@cir-innovations.ru}
\altaffiliation[ORCID: ]{0000-0003-3588-2894}
\affiliation{Joint-Stock Company ``Center for Research and Development'',
Moscow 101000, Russia}

\author{Qing-Hua Qin}
\email{qinghua.qin@smbu.edu.cn}
\altaffiliation[ORCID: ]{0000-0003-0948-784X}
\affiliation{Institute of Advanced Interdisciplinary Technology,
Shenzhen MSU-BIT University, Shenzhen 518172, China}

\date{\today}

\begin{abstract}
Carrier-envelope-phase (CEP) stabilized few-cycle pulses make the sub-cycle
field structure a control parameter for laser-driven polarized electron
sources, yet spin solvers for such pulses are rarely tested against exact
results. We show that for an electron initially at rest in a linearly
polarized plane-wave pulse the rest-frame polarization angle follows the
instantaneous vector potential, $\Sigma=2\arctan(a_{x}/2)+a_{e}a_{x}$ with
$a_{e}$ the electron anomaly: the net rotation vanishes for every CEP, while the peak intra-pulse
angle varies over the CEP by $3.02^{\circ}$ for a two-cycle and by
$0.22^{\circ}$ for an eight-cycle pulse. The result follows from the Volkov
orbit in longitudinal and transverse rapidities. We use it to validate a
light-front reference integrator and a physics-informed neural network trained
only on the light-front equations, which reaches $4\times10^{-5}$ in the spin
sector and carries over to elliptical polarization, for which no closed-form
solution is known. This gives validated tools for CEP-resolved spin dynamics in intense
fields.
\end{abstract}

\maketitle

For a plane-wave pulse of arbitrary envelope and carrier-envelope phase (CEP)
the orbit of a relativistic electron is the exact Volkov
solution~\cite{Wolkow1935}, because the light-front projection $k\cdot u$ is
conserved for any potential $A^{\mu}(\eta)$, $\eta=k\cdot x$. The
Bargmann--Michel--Telegdi (BMT) spin
equation~\cite{BMT1959,Thomas1926,Costella2001} along that orbit is then linear
in the light-front phase: at $g=2$ its propagator is a null rotation in the
little group of $k$~\cite{Kupersztych1976}, and for linear polarization the
anomalous-moment part is a rotation about a fixed axis, so that the spin is an
algebraic function of the instantaneous vector potential at any
$g$~\cite{Ternov1968,Bagrov2014,Walser2002}. For elliptical polarization at
$g\neq2$ no closed-form solution is known beyond the constant-modulus
profile~\cite{Bagrov2014}, and the net rotation after the pulse is a holonomy
of second order in the anomaly~\cite{Akintsov2026holonomy}. Beyond the classical
BMT picture, loop corrections make the spin coupling field
dependent~\cite{Ilderton2020}, and temporally asymmetric subcycle or chirped
pulses rotate the spin of relativistic beams~\cite{Wei2023}. Resolving the
sub-cycle, CEP-dependent spin dynamics inside a pulse therefore calls for
numerical solvers in the general case, and an exactly solvable configuration is
the natural place to validate them. CEP-stabilized few-cycle pulses are now routine
\cite{Paulus2001,Baltuska2003,Kienberger2004,Kling2006,Heide2019},
multi-petawatt facilities are in operation~\cite{Radier2022}, focused
intensities above $10^{23}\,\mathrm{W\,cm^{-2}}$ have been
reached~\cite{Yoon2021}, and sources of spin-polarized electron and photon
beams are being developed for laser--electron collisions and plasma
accelerators~\cite{Li2019,Li2020,Nie2021,Wen2019,Wu2020,Buescher2020}.
Predicting how the sub-cycle structure of the driving field imprints itself on
the electron polarization is therefore a prerequisite for designing
laser-driven polarized particle sources. Rapidity-based coupled-parameter
descriptions and Lorentz-constrained neural-network modeling of
spin-independent electron motion in focused Gaussian laser fields were
developed in our recent work~\cite{AkintsovPPCF2026}.

Here we show that parameterizing the four-velocity by longitudinal and
transverse rapidity variables, tied together by the light-front constant, and
the spin by its rest-frame polarization angle yields a closed light-front system for the joint orbital
and spin evolution, explicitly solvable for linear polarization. Three results
follow. First, for a linearly polarized plane-wave pulse the net spin rotation
accumulated over the complete pulse vanishes identically for any CEP and any
$g$~\cite{Ternov1968,Bagrov2014,Walser2002}; in the rapidity formulation this
appears as a total derivative in the light-front phase, and direct integration
reproduces it to $10^{-12}$ degrees without the invariant ever being imposed.
Second, the physically accessible CEP effect is therefore not a net flip but a
transient intra-pulse rotation of the rest-frame polarization, fixed at $g=2$
by the instantaneous vector potential and hence strongly CEP dependent in the
few-cycle limit and essentially CEP independent for longer pulses. Third, the
solvers are validated: for a co-propagating electron the
light-front parameterization converges at the same fourth order as lab-time
integration but reaches a roundoff floor an order of magnitude lower, and a
physics-informed neural network (PINN) trained purely on the light-front
orbital and spin equations reproduces the exact solution without labeled data
and carries over to elliptical polarization, for which no closed-form solution
is known.

What is new relative to the exact solutions of
Refs.~\cite{Ternov1968,Bagrov2014,Walser2002}, to the holonomy of
Ref.~\cite{Akintsov2026holonomy} and to the spin-independent Gaussian-field
modeling of Ref.~\cite{AkintsovPPCF2026} is (i) the explicit intra-pulse polarization
angle and its equation of motion in rapidity variables, and the CEP dependence
they imply for few-cycle pulses; (ii) a fixed-step-budget comparison of light-front and lab-time
integrators for a co-propagating electron, including the structure-preserving
Higuera--Cary pusher~\cite{HigueraCary2017}; and (iii) a physics-informed solver for the
coupled orbital and spin equations, validated against that benchmark with a
controlled ablation of its input embedding and applied to a case for which no
closed-form solution is known.

\emph{Rapidity formulation.}---We parameterize the four-velocity by two
rapidity variables---a longitudinal (light-front) rapidity $\theta$ and a
transverse rapidity $\phi$---rather than by the velocity components,
\begin{equation}
\begin{split}
u^{\mu}&=\big[e^{\theta K_{z}}e^{\phi K_{x}}\big]^{\mu}{}_{0}\\
&=(\cosh\theta\cosh\phi,\,\sinh\phi,\,0,\,\cosh\phi\sinh\theta),
\end{split}
\label{eq:lorentz}
\end{equation}
in units $c=m=1$, where $K_{x}$ and $K_{z}$ generate boosts along $x$ and $z$,
$u\cdot u=1$ holds identically, and the energy is
$\gamma=\cosh\theta\cosh\phi$. The pulse propagates along $+z$ and is polarized
along $x$, with light-front phase $\eta=\omega(t-z)$ and normalized potential
$a_{x}(\eta)=|e|A_{x}/m$. The pair $(\theta,\phi)$ is not independent: the
light-front constant of the plane-wave motion,
$\kappa\equiv k\!\cdot\!u/\omega=\gamma-u_{z}=\cosh\phi\,e^{-\theta}$, ties
their evolution together, and the orbit is
\begin{equation}
\sinh\phi=u_{x0}+a_{x}(\eta)-a_{x}(0),\qquad
\theta=\ln(\cosh\phi/\kappa),
\label{eq:ham}
\end{equation}
the Volkov solution written in rapidities. It holds to all orders in the field
strength, requires none of the small-parameter expansions that limit
perturbative treatments, and reduces to a single longitudinal boost for
$\phi\to0$. The derivation of Eqs.~(\ref{eq:lorentz})--(\ref{eq:ham}) and of
the spin equations below is given in Secs.~\ref{sec:s1} and~\ref{sec:s2} of the
Supplemental Material.

\emph{Rapidity--BMT spin equations.}---The spin four-vector $S^{\mu}$,
$S\cdot u=0$, $S\cdot S=-1$, obeys the covariant BMT equation
\begin{equation}
\frac{\ud S^{\mu}}{\ud\tau}
=\frac{q}{m}\Big[\frac{g}{2}\,F^{\mu\nu}S_{\nu}
+\Big(\frac{g}{2}-1\Big)u^{\mu}\big(S_{\lambda}F^{\lambda\nu}u_{\nu}\big)\Big],
\label{eq:bmt}
\end{equation}
with $q=-|e|$ the electron charge, $F^{\mu\nu}$ the field tensor and
$g=2(1+a_{e})$ the electron gyromagnetic ratio. The orbital block does not
depend on the spin and, for a plane-wave pulse of any envelope and CEP, is
solved algebraically by Eq.~(\ref{eq:ham}); Eq.~(\ref{eq:bmt}) is then a linear
equation for $S^{\mu}$ with coefficients that are known functions of
$\eta$ (Sec.~\ref{sec:s1}). Writing the spin as
$S^{\mu}=L^{\mu}{}_{\nu}(u)\,(0,\bm{\zeta})^{\nu}$, with $L(u)$ the pure boost
to the instantaneous rest frame and $\bm{\zeta}=(-\sin\Sigma,0,\cos\Sigma)$ the
rest-frame polarization at angle $\Sigma$ from the propagation axis, the system
closes in the light-front phase. For an electron initially at rest,
\begin{align}
\frac{\ud\Sigma}{\ud\eta}
&=\Omega_{0}+\Big(\frac{g}{2}-1\Big)\Omega_{a},\quad
\Omega_{0}=\frac{a_{x}'}{1+a_{x}^{2}/4},\quad \Omega_{a}=a_{x}',
\label{eq:sigma}\\[2pt]
\frac{\ud\phi}{\ud\eta}&=\frac{a_{x}'}{\cosh\phi},\qquad
\frac{\ud\theta}{\ud\eta}=\tanh\phi\,\frac{\ud\phi}{\ud\eta},
\label{eq:rapidity}
\end{align}
where $a_{x}'=\ud a_{x}/\ud\eta$, $\Omega_{0}$ is the precession of a $g=2$
moment, which reduces to the rest-frame Larmor precession in the magnetic field
of the wave for $|a_{x}|\ll1$ and carries its relativistic (Thomas) corrections
at finite $a_{x}$, and $\Omega_{a}$ is the anomalous precession. Both rotate
$\bm{\zeta}$ about
$\hat{\bm y}$, the normal to the plane of the orbit, so
Eqs.~(\ref{eq:sigma})--(\ref{eq:rapidity}) form a closed first-order system
that requires no separate integration of the four-velocity and integrates in
closed form, $\Sigma(\eta)=2\arctan[a_{x}(\eta)/2]+a_{e}a_{x}(\eta)$. For
elliptical polarization the axis of $\Omega_{a}$ turns with
$\bm{a}_{\perp}'$, the rotations at different phases no longer commute, and for
$g\neq2$ the closed form is lost~\cite{Bagrov2014,Akintsov2026holonomy}; that is the regime
the solvers below are built for, and the linear case is their exact
benchmark.

\emph{Carrier-envelope-phase-driven spin dynamics.}---We take a linearly
polarized pulse
$a_{x}(\eta)=a_{0}\,\mathcal{E}(\eta)\cos(\eta+\CEP)$ with a $\cos^{2}$
envelope $\mathcal{E}$ of $N$ optical cycles, $a_{0}=0.42$, and an electron
initially at rest with its spin along the propagation axis
$S^{\mu}(0)=(0,0,0,1)$. Reference solutions are obtained by integrating
Eq.~(\ref{eq:bmt}) in $\eta$ along the orbit of Eq.~(\ref{eq:ham}) with an
eighth-order Dormand--Prince integrator at relative and absolute tolerances
$10^{-11}$ and $10^{-13}$; they coincide with the closed form of
Eq.~(\ref{eq:sigma}) to $10^{-12}$ at every phase, and parameters and
convergence tests are listed in Secs.~\ref{sec:s3} and~\ref{sec:s7}.

Equation~(\ref{eq:sigma}) makes an exact statement possible before any
numerics. For a linearly polarized pulse of compact support both terms are
total derivatives of functions of $a_{x}(\eta)$, so that
\begin{equation}
\Delta\Sigma\big|_{\eta\ge2\pi N}
=\Big[2\arctan\frac{a_{x}}{2}+a_{e}a_{x}\Big]_{0}^{2\pi N}=0
\quad \text{for all } \CEP ,
\label{eq:invariant}
\end{equation}
i.e.\ the electron leaves the pulse with exactly the polarization it entered
with, at any $g$ and any $a_{0}$~\cite{Ternov1968,Bagrov2014,Walser2002}.
Numerically, scanning 16 values of the CEP over $[0,2\pi)$ we find a residual
net rotation of at most $3\times10^{-13}$ degrees for $N=2$ and
$9\times10^{-13}$ degrees for $N=8$, at the level of the integrator tolerance.
For elliptical polarization the net rotation is instead a holonomy
of magnitude $\tfrac12a_{e}^{2}|\mathcal{A}|$ about the propagation axis, with
$\mathcal{A}$ twice the signed area enclosed by $\bm{a}_{\perp}(\eta)$ in the
polarization plane; it is nonzero, CEP independent and of second order in the
anomaly~\cite{Akintsov2026holonomy}, and for circular polarization our
integrator reproduces it to five significant figures (Sec.~\ref{sec:s1ellip}). This is a
constraint rather than a null result: any reported CEP-driven \emph{net}
polarization change for a plane-wave pulse must originate from an ingredient
outside the ideal plane-wave BMT description---finite focusing, radiative
corrections, or a scattering event occurring while the pulse is still on.

\begin{figure}[t]
\centering
\includegraphics[width=\columnwidth]{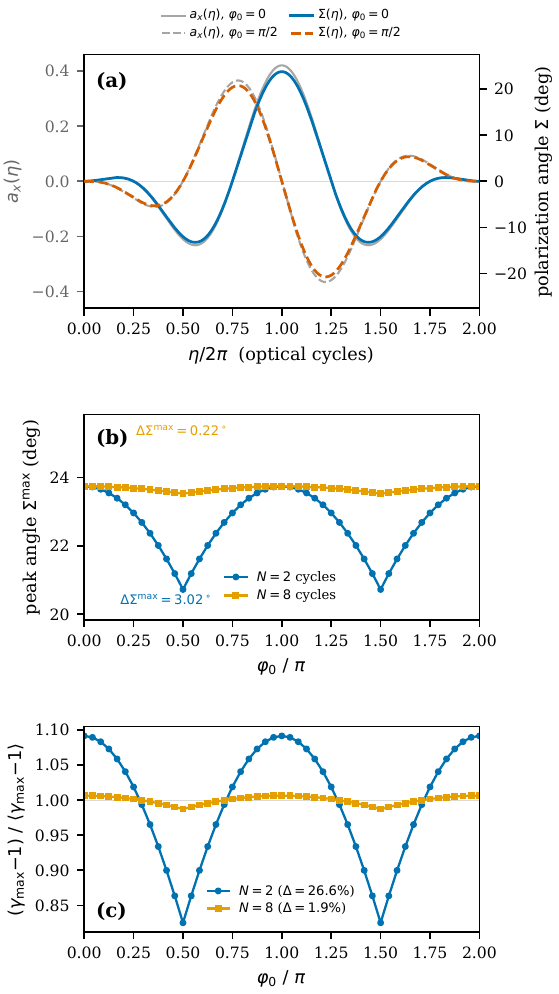}
\caption{Carrier-envelope-phase control of the intra-pulse spin dynamics for
an electron initially at rest in a $\cos^{2}$-envelope pulse with $a_{0}=0.42$.
(a) Normalized vector potential $a_{x}(\eta)$ (grey) and the rest-frame
polarization angle $\Sigma(\eta)$ for $\CEP=0$ and $\CEP=\pi/2$ in a two-cycle
pulse; $\Sigma=2\arctan(a_{x}/2)+a_{e}a_{x}$.
(b) Peak angle $\Sigma^{\max}=\max_{\eta}|\Sigma|$ versus CEP for $N=2$ and
$N=8$ cycles; the CEP-induced spread collapses from $3.02^{\circ}$ to
$0.22^{\circ}$ as the pulse lengthens.
(c) Peak kinetic energy $\gamma_{\max}-1$, normalized to its CEP average,
showing a $26.6\%$ spread for $N=2$ against $1.90\%$ for $N=8$;
$\gamma_{\max}-1=a_{\max}^{2}/2$ exactly, with $a_{\max}=\max_{\eta}|a_{x}|$.
In all cases the \emph{net} rotation after the pulse vanishes,
Eq.~(\ref{eq:invariant}).}
\label{fig:cep}
\end{figure}

What the CEP does control is the transient. Figure~\ref{fig:cep}(a) shows
$\Sigma(\eta)$ for two CEP values in a two-cycle pulse: the polarization follows
the instantaneous vector potential, reaching up to $23.7^{\circ}$ at the peak of the
potential, where the electric field vanishes, and returns to its initial
direction. Hence $\Sigma^{\max}=2\arctan(a_{\max}/2)+a_{e}a_{\max}$ with
$a_{\max}=\max_{\eta}|a_{x}|$; the anomalous moment contributes
$a_{e}a_{\max}\simeq0.03^{\circ}$, so the excursion is kinematic. Scanning the
CEP [Fig.~\ref{fig:cep}(b)] gives a spread of $\Sigma^{\max}$ of
$3.02^{\circ}$ for $N=2$ and only $0.22^{\circ}$ for $N=8$, a suppression by a
factor $13.8$ that is the CEP dependence of $a_{\max}$
($0.365$--$0.420$ for $N=2$, $0.416$--$0.420$ for $N=8$), i.e.\ of the weight of the
envelope gradient $\ud\mathcal{E}/\ud\eta$ relative to the carrier. The
orbital sector behaves the same way: the peak kinetic energy varies by
$26.6\%$ over the CEP for the two-cycle pulse and by $1.90\%$ for the
eight-cycle pulse [Fig.~\ref{fig:cep}(c)], and these relative spreads are independent of
$a_{0}$ identically, since $\gamma_{\max}-1=a_{\max}^{2}/2$ with
$a_{\max}\propto a_{0}$. The intra-pulse polarization is the quantity to which
any sub-cycle probe couples---Compton or Thomson scattering off the electron
while the pulse is still present, or a truncated interaction---so
Fig.~\ref{fig:cep}(b) is directly the CEP-tagged observable for such a
measurement, in the same spirit as CEP-resolved nonlinear Compton
scattering~\cite{Mackenroth2010,Krajewska2012}.

\begin{figure}[tp]
\centering
\includegraphics[width=0.95\columnwidth]{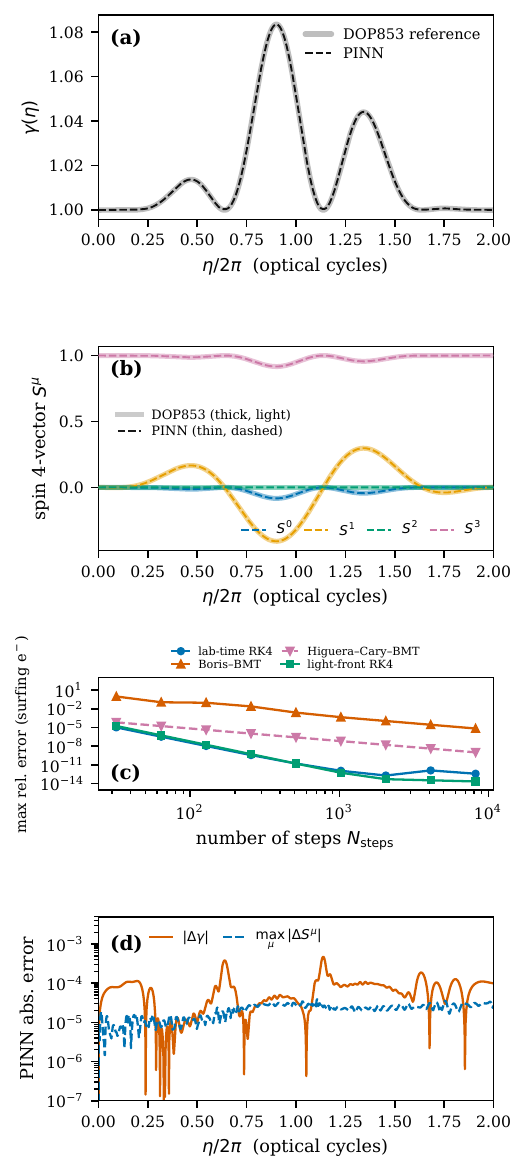}
\caption{Physics-informed neural network solution of the light-front orbital
and BMT equations, Eq.~(\ref{eq:bmt}) with $\ud/\ud\tau=\kappa\,\ud/\ud\eta$,
for
$a_{0}=0.42$, $N=2$, $\CEP=0.7$~rad.
(a) Lorentz factor $\gamma(\eta)$ from the PINN (dashed) against the
Dormand--Prince reference (solid).
(b) Components of the spin four-vector $S^{\mu}(\eta)$, same convention.
(c) Maximum relative error at the end of the pulse versus fixed step budget for
a co-propagating electron ($\kappa=0.050$), comparing lab-time Runge--Kutta,
the Boris and Higuera--Cary pushers (both with a Cayley spin update) and the
light-front rapidity parameterization.
(d) Pointwise absolute error of the trained network against the reference,
$|\Delta\gamma|$ and $\max_{\mu}|\Delta S^{\mu}|$, over the pulse.}
\label{fig:pinn}
\end{figure}

\emph{Light-front and neural solvers.}---For a co-propagating electron the
lab-time formulation loses precision. Taking $\gamma_{0}=10$, so that
$\kappa=\gamma_{0}-u_{z,0}=0.050$, and integrating with a fixed step budget, we
find [Fig.~\ref{fig:pinn}(c)] that lab-time Runge--Kutta and the light-front
parameterization, which integrates only the four spin equations along the exact
orbit, converge at the same fourth order, and that lab-time Runge--Kutta then
stalls at a roundoff floor near $10^{-12}$ while the light-front scheme
continues to $2\times10^{-14}$, because the phase $\eta=\omega(t-z)$ is never
formed by subtracting two large numbers. The Boris~\cite{Boris1970} and
Higuera--Cary~\cite{HigueraCary2017} pushers---the latter removing, like the
Vay pusher~\cite{Vay2008}, the spurious force of the Boris rotation on a
relativistic particle---both converge at second order, to
$3\times10^{-5}$ and $4\times10^{-9}$ at 4096 steps.

The light-front orbital and BMT equations, Eq.~(\ref{eq:bmt}) with
$\ud/\ud\tau=\kappa\,\ud/\ud\eta$, are also a natural target for a
physics-informed neural network~\cite{Raissi2019}, and the closed form of
Eq.~(\ref{eq:sigma}) provides an exact benchmark for it. A related
Lorentz-constrained network was applied to spin-independent motion in focused
Gaussian fields~\cite{AkintsovPPCF2026}; the network here solves the coupled
orbital and spin equations and is trained without labeled trajectory data. We
represent the state
by a feedforward network $N_{\Theta}(\eta)=(\gamma,\uv,S^{\mu})$ with a
sinusoidal (Fourier-feature) input embedding of the light-front phase, which
counters the spectral bias of neural
networks~\cite{Rahaman2019,Tancik2020,Wang2021}, and hard-wired initial
conditions, and minimize
\begin{equation}
\mathcal{L}(\Theta)=\mathcal{L}_{\mathrm{orb}}
+\lambda_{s}\mathcal{L}_{\mathrm{spin}}
+\lambda_{i}\big(\mathcal{L}_{u\cdot u}+\mathcal{L}_{S\cdot S}\big),
\label{eq:loss}
\end{equation}
where $\mathcal{L}_{\mathrm{orb}}$ and $\mathcal{L}_{\mathrm{spin}}$ are the
mean-squared residuals of the orbital and BMT blocks evaluated by automatic
differentiation, and the last two terms penalize violation of the mass-shell
condition $u\!\cdot\!u=1$ and the spin normalization $S\!\cdot\!S=-1$. No
labeled trajectory data enter the loss at any point: the network is trained
only on the equations and their invariants. Architecture, collocation
sampling, loss weighting and the optimizer schedule are given
in Sec.~\ref{sec:s6}.

% Four stacked panels: too tall to share a column with text, so the figure is
% placed on a float page of its own; at full column width it also overruns the
% page height by ~17pt, hence the 0.95 scale.

Figure~\ref{fig:pinn}(a)--(b) compares the trained network against the
Dormand--Prince reference over the full pulse. The trained network reaches a
maximum absolute error of $4.7\times10^{-4}$ in $\gamma(\eta)$ and $3.9\times10^{-5}$ in the
spin components [Fig.~\ref{fig:pinn}(d)], the same against the exact solution,
while the invariant term of Eq.~(\ref{eq:loss}) is driven to $2\times10^{-8}$, so that
$u\!\cdot\!u$ and $S\!\cdot\!S$ are preserved to $9\times10^{-4}$ at every point of
the pulse without either being imposed algebraically (Sec.~\ref{sec:s6acc}). The
embedding matters mainly in the spin sector: the same network, trained with the
same schedule and optimizer but supplied only with the phase, reaches
$8\times10^{-4}$ in $\gamma$ and $4\times10^{-4}$ in the spin components, larger
by factors of $1.7$ and $10$; over two carrier
cycles the spectral bias of physics-informed
networks~\cite{Krishnapriyan2021,Wang2021} slows convergence rather than
preventing it. For an elliptically polarized pulse
(ellipticity $0.5$, same $a_{0}$, $N$ and $\CEP$), for which no closed-form solution is known,
the same network reaches $9\times10^{-5}$ in $\gamma$ and $2\times10^{-5}$ in the spin
components against the reference (Sec.~\ref{sec:s6ell}). Each network is trained for
one parameter point (74--87~min on a CPU), whereas a Dormand--Prince
solution takes a fraction of a second; the network provides a mesh-free,
differentiable representation of the orbit and spin checked against an exact
result, and a parametric extension in $(\CEP,N)$ is left to future work.

\emph{Implications and outlook.}---Two conclusions follow for laser-driven
polarized electron sources. The exact invariant, Eq.~(\ref{eq:invariant}),
shows that CEP alone cannot switch the polarization of an electron that
traverses a linearly polarized plane-wave pulse in vacuum, and for elliptical
polarization the only net rotation is the CEP-independent holonomy of
Ref.~\cite{Akintsov2026holonomy}; proposals for CEP-controlled polarization
must therefore be built on the mechanisms that break the plane-wave
idealization. Conversely, the strong CEP dependence of the intra-pulse
polarization, $3.02^{\circ}$ against $0.22^{\circ}$ between two- and eight-cycle
pulses, makes the transient spin state a CEP-tagged record of the peak vector
potential inside the pulse, accessible to any probe that couples to the
electron while the pulse is on.

Because the network returns a continuously differentiable surrogate of the
single-particle orbital and spin dynamics, it can serve as a microscopic input
for kinetic or spin-hydrodynamic descriptions of polarized
ensembles~\cite{Florkowski2019,Bhadury2021}; we have not pursued this here.

\begin{acknowledgments}
Earlier versions of this work were presented at the 26th International
Symposium on Spin Physics (SPIN2025), Qingdao, China, 21--26 September
2025~\cite{AkintsovSPIN2025}, and at the 1st Conference on Strong-Interaction
Spin Physics, Qingdao, China, 26--31 July 2026~\cite{AkintsovSISP2026}. The
authors thank their colleagues at Nantong University and Kuban State University
for discussions.

This work was partially supported by the State Assignment of the Ministry of
Education and Science of the Russian Federation (Project No.\ FZEN 2023-0006),
the Nantong Science and Technology Plan Project (Grant Nos.\ JC2020137 and
JC2020138), the Key Research and Development Program of Jiangsu Province of
China (Grant No.\ BE2021013-1), the National Natural Science Foundation of
Jiangsu Province of China (Grant No.\ BK20201438), and in part by the Natural
Science Research Project of Jiangsu Provincial Institutions of Higher Education
(Grant Nos.\ 1120KJA510002 and 20KJB510010).
\end{acknowledgments}

\section*{Data availability}
The code that generates the reference solutions, the convergence study, the
neural-network training and all figures of this Letter, together with the
computed data and the trained networks, is openly available in the Zenodo
repository at Ref.~\cite{Akintsov2026code} (release 1.0.1).

\bibliographystyle{apsrev4-2}
\bibliography{refs}

% ----------------------------------------------------------------------
%  Supplemental Material, appended for the preprint. In the journal
%  submission this is a separate file, pra_submission_v2/supplement.tex.
%  It is set one-column: its displayed equations and tables are typeset
%  for the full page width.
% ----------------------------------------------------------------------
\clearpage
\onecolumngrid

\setcounter{section}{0}
\setcounter{equation}{0}
\setcounter{table}{0}
\setcounter{figure}{0}
\renewcommand{\thesection}{S\arabic{section}}
\renewcommand{\thesubsection}{S\arabic{section}\,\Alph{subsection}}
% \thesubsection already carries the section number, so the \ref prefix
% \p@subsection (= \thesection here) must be cleared, or references to a
% subsection come out as "Sec. S1 S1 D".
\makeatletter
\renewcommand{\p@subsection}{}
\makeatother
\renewcommand{\theequation}{S\arabic{equation}}
\renewcommand{\thetable}{S\arabic{table}}
\renewcommand{\thefigure}{S\arabic{figure}}

\begin{center}
{\large\bfseries Supplemental Material}
\end{center}

This Supplemental Material provides the exact plane-wave reduction on which
the Letter rests, with the closed-form solution for linear polarization and the
scope of closed-form solutions (Sec.~\ref{sec:s1}); the rapidity
parameterization and the derivation of the rapidity--BMT equations of the main
text (Sec.~\ref{sec:s2}); the fixed-step-budget convergence study comparing
lab-time Runge--Kutta, the Boris and Higuera--Cary pushers and the light-front
parameterization (Sec.~\ref{sec:s3}); the numerical protocol used to extract the
CEP-controlled intra-pulse polarization reported in the main text
(Sec.~\ref{sec:s5}); and the architecture, training protocol, controlled
ablation and elliptically polarized test of the physics-informed neural network
(PINN) solver (Sec.~\ref{sec:s6}), together with a reproducibility summary
(Sec.~\ref{sec:s7}).

\section{Exact plane-wave reduction and closed-form solutions}
\label{sec:s1}

\subsection{Volkov orbit for an arbitrary pulse}

We use units $c=m=|e|=1$, the metric $(+,-,-,-)$ and the electron charge
$q=-1$. For a potential that depends on the light-front phase alone,
$\bm{a}_{\perp}=\bm{a}_{\perp}(\eta)$ with $\eta=k\cdot x=\omega(t-z)$ and
$k^{2}=0$, the field tensor is built from $k^{\mu}$ and
$\bm{a}_{\perp}'=\ud\bm{a}_{\perp}/\ud\eta$ alone, and $k_{\mu}F^{\mu\nu}=0$
identically, whatever the envelope and the carrier-envelope phase. The
light-front constant $\kappa=k\cdot u/\omega=\gamma-u_{z}$ is therefore
conserved, $\ud\eta/\ud\tau=\omega\kappa$, and so is the transverse canonical
momentum $\bm{u}_{\perp}-\bm{a}_{\perp}(\eta)$. The orbit is algebraic in
$\eta$,
\begin{equation}
\bm{u}_{\perp}=\bm{u}_{\perp0}+\bm{a}_{\perp}(\eta)-\bm{a}_{\perp}(0),\qquad
\gamma=\frac{1}{2}\Big[\kappa+\frac{1+\bm{u}_{\perp}^{2}}{\kappa}\Big],\qquad
u_{z}=\gamma-\kappa,
\label{eq:volkov}
\end{equation}
the classical counterpart of the Volkov solution~\cite{Wolkow1935}. A pulse
envelope does not break this structure, because for a pulse propagating in
vacuum it is itself a function of $\eta$. For an electron initially at rest
$\kappa=1$ and $\gamma-1=a_{x}^{2}/2$ exactly. The reference integrator, the
convergence study and the PINN of the main text all rest on
Eq.~(\ref{eq:volkov}).

\subsection{Spin: the $g=2$ propagator and the anomalous part}

Along the orbit (\ref{eq:volkov}) the BMT equation [main text Eq.~(3)] is a
linear equation for $S^{\mu}$ with coefficients that are known functions of
$\eta$. At $g=2$ its generator is a combination of the two null rotations
$N_{i}=k\wedge e_{i}$ of the little group of $k$, which commute and are
nilpotent, so the propagator is the null rotation
$\Lambda_{0}(\eta)=\exp[\kappa^{-1}(a_{1}N_{1}+a_{2}N_{2})]$ for any
pulse~\cite{Kupersztych1976}; it is the same Lorentz transformation that
carries $u(0)$ to $u(\eta)$. Factoring the propagator as $\Lambda_{0}V$, the
remaining evolution $V$ is a rotation in the rest frame of $u(0)$ driven by the
anomaly alone, about the axis $\hat{\bm n}\times\bm{a}_{\perp}'$, with
$\hat{\bm n}$ the propagation
direction~\cite{Ternov1968,Bagrov2014,Akintsov2026holonomy}.

\subsection{Closed form for linear polarization}

For linear polarization along $x$ that axis is fixed along $\hat{\bm y}$, the
rotations at different phases commute, and $V$ is a rotation through the angle
$a_{e}[a_{x}(\eta)-a_{x}(0)]$~\cite{Bagrov2014,Walser2002}. For an electron
initially at rest with $S^{\mu}(0)=(0,0,0,1)$ this gives, with $a=a_{x}(\eta)$,
\begin{equation}
S^{\mu}(\eta)=\Big(\zeta_{x}a-\zeta_{z}\frac{a^{2}}{2},\ \zeta_{x}-\zeta_{z}a,\ 0,\
\zeta_{x}a+\zeta_{z}\Big(1-\frac{a^{2}}{2}\Big)\Big),\quad
\zeta_{x}=-\sin(a_{e}a),\ \ \zeta_{z}=\cos(a_{e}a),
\label{eq:closedS}
\end{equation}
at any $g$, envelope, CEP and $a_{0}$. Equation~(\ref{eq:closedS}) coincides
with the DOP853 reference of Sec.~\ref{sec:s5} to $8\times10^{-13}$ for the physical
anomaly and for anomalies inflated to $a_{e}=0.05$ and $0.3$, for $N=2$ and $8$
and several CEP values. Two consequences are used in the main text. At the end
of the pulse $a_{x}=0$, so $S^{\mu}$ returns to its initial value exactly, for
every CEP and every $g$. And the rest-frame polarization follows the
instantaneous vector potential (Sec.~\ref{sec:s2}).

\subsection{Elliptical polarization}
\label{sec:s1ellip}

For elliptical polarization at $g\neq2$ the axis
$\hat{\bm n}\times\bm{a}_{\perp}'$ turns with the potential, the rotations at
different phases do not commute, and beyond linear polarization a closed form is
known only for the constant-modulus profile~\cite{Bagrov2014}. The net rotation
after the pulse is then a holonomy of magnitude $\tfrac12a_{e}^{2}|\mathcal{A}|$
about the propagation axis, with $\mathcal{A}=\int(a_{x}a_{y}'-a_{y}a_{x}')\,\ud\eta$
twice the signed area enclosed by the curve
$\bm{a}_{\perp}(\eta)$~\cite{Akintsov2026holonomy}. For a circularly polarized
$\cos^{2}$ pulse with $a_{0}=0.42$ (peak $|\bm{a}_{\perp}|=a_{0}/\sqrt{2}$) and a
transverse initial spin our integrator gives net rotations of $2.7947\times10^{-7}~\mathrm{rad}$ for $N=2$
and $1.1179\times10^{-6}~\mathrm{rad}$ for $N=8$, equal to $\tfrac12a_{e}^{2}|\mathcal{A}|$ to five
significant figures and identical for all CEP values tested. The vanishing of
the net rotation used in the main text, Eq.~(6), is therefore a property of
linear polarization. Beyond the classical BMT description, the spin of an
electron in counterpropagating laser beams can also be rotated through its
coupling to the photon spin~\cite{Bauke2014}.

\section{Rapidity parameterization and the rapidity--BMT equations}
\label{sec:s2}

\subsection{Rapidities and the orbit}

We write the four-velocity as the image of the rest-frame vector under an
ordered product of two boosts,
\begin{equation}
u^{\mu}=\big[e^{\theta K_{z}}e^{\phi K_{x}}\big]^{\mu}{}_{\nu}\,(1,0,0,0)^{\nu}
=(\cosh\theta\cosh\phi,\ \sinh\phi,\ 0,\ \cosh\phi\sinh\theta),
\label{eq:urap}
\end{equation}
where $K_{x}$ and $K_{z}$ generate boosts along $x$ and $z$. The two generators
do not commute, and the order of the factors is part of the definition. The
mass shell is satisfied identically,
$u\cdot u=\cosh^{2}\phi\,(\cosh^{2}\theta-\sinh^{2}\theta)-\sinh^{2}\phi=1$, the
energy is $\gamma=\cosh\theta\cosh\phi$, and the light-front constant is
$\kappa=\gamma-u_{z}=\cosh\phi\,e^{-\theta}$. Inserting Eq.~(\ref{eq:urap}) in
Eq.~(\ref{eq:volkov}) for a pulse polarized along $x$ gives the orbit of main
text Eq.~(2), $\sinh\phi=u_{x0}+a_{x}(\eta)-a_{x}(0)$ and
$\theta=\ln(\cosh\phi/\kappa)$; differentiating at constant $\kappa$ gives main
text Eq.~(5), $\ud\phi/\ud\eta=a_{x}'/\cosh\phi$ and
$\ud\theta/\ud\eta=\tanh\phi\,\ud\phi/\ud\eta$. Rapidity-based
coupled-parameter descriptions of relativistic charged-particle motion were
developed in our earlier work~\cite{Akintsov2023,Akintsov2024,Akintsov2025} and
recently extended to spin-independent electron dynamics in focused Gaussian
laser fields with Lorentz-constrained neural-network
modeling~\cite{AkintsovPPCF2026}. The construction here uses two ordered
rapidities tied by the light-front invariant and carries the resulting system
over to the covariant BMT spin dynamics.

\subsection{Rest-frame polarization and the $g=2$ precession}

The rest-frame polarization $\bm{\zeta}$ is defined with the pure boost $L(u)$
to the instantaneous rest frame, $S^{\mu}=L^{\mu}{}_{\nu}(u)\,(0,\bm{\zeta})^{\nu}$,
equivalently $\bm{\zeta}=\bm{S}-S^{0}\bm{u}/(\gamma+1)$. For an electron
initially at rest and $g=2$ one has $\kappa=1$, $u=(1+a^{2}/2,\,a,\,0,\,a^{2}/2)$
and, from Eq.~(\ref{eq:closedS}), $S=(-a^{2}/2,\,-a,\,0,\,1-a^{2}/2)$ with
$a=a_{x}(\eta)$, whence
\begin{equation}
\bm{\zeta}=\Big(-\frac{4a}{4+a^{2}},\ 0,\ \frac{4-a^{2}}{4+a^{2}}\Big)
=(-\sin\Sigma_{0},\,0,\,\cos\Sigma_{0}),\qquad
\Sigma_{0}=2\arctan\frac{a_{x}}{2}.
\label{eq:sigmaT}
\end{equation}
The null rotation $\Lambda_{0}$ is thus the pure boost $L(u)$ composed with a
rotation about $\hat{\bm y}$ through $\Sigma_{0}$, which is the whole spin
precession of a $g=2$ moment along the orbit: for $|a_{x}|\ll1$,
$\ud\Sigma_{0}/\ud\eta\simeq a_{x}'$ is the rest-frame Larmor precession in the
magnetic field of the wave, and the factor $(1+a_{x}^{2}/4)^{-1}$ carries its
relativistic (Thomas) corrections~\cite{Thomas1926}. The anomalous rotation of
Sec.~\ref{sec:s1} acts about the same axis and in the same sense, so the angles
add,
\begin{equation}
\Sigma(\eta)=2\arctan\frac{a_{x}(\eta)}{2}+a_{e}a_{x}(\eta),\qquad
\frac{\ud\Sigma}{\ud\eta}=\frac{a_{x}'}{1+a_{x}^{2}/4}+a_{e}a_{x}',
\label{eq:sigmafull}
\end{equation}
which is main text Eq.~(4) with $\Omega_{0}=a_{x}'/(1+a_{x}^{2}/4)$ and
$\Omega_{a}=a_{x}'$. The angle $\Sigma$ extracted from the DOP853 reference
agrees with Eq.~(\ref{eq:sigmafull}) to $4\times10^{-13}~\mathrm{rad}$, including for inflated
anomalies, which fixes the relative sense of the two rotations. The angle
$\theta_{S}$ between the spatial part of the laboratory four-vector $S^{\mu}$
and the propagation axis is frame dependent and is not a polarization; it
differs from $\Sigma$ by the boost: at $|a_{x}|=0.42$ one has
$\theta_{S}=24.73^{\circ}$ at $g=2$, whereas $\Sigma_{0}=23.72^{\circ}$ and
the anomalous term adds $0.03^{\circ}$.

\subsection{Scope}

Equations~(4)--(5) of the main text are written for linear polarization and an
electron initially at rest, the configuration of the CEP scans and of the main
PINN benchmark. For elliptical polarization the anomalous rotation no longer
shares the axis of the $g=2$ precession and Eq.~(\ref{eq:sigmafull}) does not
apply (Sec.~\ref{sec:s1ellip}); the PINN is tested in that case in
Sec.~\ref{sec:s6ell}.

\section{Fixed-step-budget convergence study}
\label{sec:s3}
\label{sec:convergence}

The convergence comparison is carried out for a co-propagating (``surfing'')
electron with a large initial forward drift, $\gamma_{0}=10$,
$u_{z,0}=\sqrt{\gamma_{0}^{2}-1}\approx9.9499$, so that the light-front constant
$\kappa=\gamma_{0}-u_{z,0}=0.050$ is small. The pulse parameters are
$a_{0}=0.42$, $N=2$ optical cycles and $\varphi_{0}=0.7$~rad, and the spin
starts longitudinal, $S^{\mu}(0)=(u_{z,0},0,0,\gamma_{0})$, which satisfies
$S\cdot u=0$ and $S\cdot S=-1$. The lab-time crossing duration is
$t_{1}=2589.69$ (units $c=m_{e}=|e|=1$, $\omega=1$). The reference is the
light-front DOP853 solution at relative and absolute tolerances $10^{-13}$ and
$10^{-15}$, and the error metric is
$\max\big(|\Delta\gamma|/\gamma_{\text{ref}},\
\max_{\mu}|\Delta S^{\mu}|/\max_{\mu}|S^{\mu}_{\text{ref}}|\big)$ at the end of
the pulse. Four fixed-step methods are compared at identical step counts
$N_{\text{steps}}$: (i) a lab-time fourth-order Runge--Kutta (RK4) integration
of the Lorentz-force and BMT equations (ten coupled equations); (ii) the
Boris~\cite{Boris1970} and (iii) the Higuera--Cary~\cite{HigueraCary2017}
pushers in lab-frame Cartesian coordinates, with positions at integer and
velocity and spin at half-integer steps, the velocity advanced by the respective
magnetic rotation and the spin by the Cayley transform of the BMT generator
evaluated at the integer step, which is second order and preserves $S\cdot S$
to roundoff; and (iv) an RK4 integration of the four light-front spin equations
along the algebraic orbit, Eq.~(\ref{eq:volkov}). Table~\ref{tab:convergence}
reports the errors.

\begin{table*}[!ht]
\caption{Maximum relative error at fixed step budget for a surfing electron
($\gamma_{0}=10$, $\kappa=0.050$, longitudinal initial spin) driven by a
$\cos^{2}$-envelope pulse with $a_{0}=0.42$, $N=2$ cycles,
$\varphi_{0}=0.7$~rad, relative to the DOP853 light-front reference.}
\label{tab:convergence}
\begin{ruledtabular}
\begin{tabular}{rcccc}
$N_{\text{steps}}$ & Lab-time RK4 & Boris--BMT & Higuera--Cary--BMT & Light-front RK4 \\
\hline
32    & $1.0\times10^{-5}$ & $9.1\times10^{-1}$ & $6.6\times10^{-5}$ & $1.7\times10^{-5}$ \\
64    & $3.3\times10^{-7}$ & $1.1\times10^{-1}$ & $1.7\times10^{-5}$ & $5.4\times10^{-7}$ \\
128   & $1.1\times10^{-8}$ & $9.0\times10^{-2}$ & $4.2\times10^{-6}$ & $1.7\times10^{-8}$ \\
256   & $3.7\times10^{-10}$ & $2.3\times10^{-2}$ & $1.0\times10^{-6}$ & $5.3\times10^{-10}$ \\
512   & $1.7\times10^{-11}$ & $2.4\times10^{-3}$ & $2.6\times10^{-7}$ & $1.6\times10^{-11}$ \\
1024  & $1.1\times10^{-12}$ & $4.4\times10^{-4}$ & $6.5\times10^{-8}$ & $5.6\times10^{-13}$ \\
2048  & $2.0\times10^{-13}$ & $1.1\times10^{-4}$ & $1.6\times10^{-8}$ & $5.2\times10^{-14}$ \\
4096  & $1.3\times10^{-12}$ & $2.7\times10^{-5}$ & $4.1\times10^{-9}$ & $3.4\times10^{-14}$ \\
8192  & $3.9\times10^{-13}$ & $6.7\times10^{-6}$ & $1.0\times10^{-9}$ & $2.2\times10^{-14}$ \\
\end{tabular}
\end{ruledtabular}
\end{table*}

Lab-time RK4 and light-front RK4 converge at the same fourth order, their errors
staying within a factor 1.7 of each other down to $2\times10^{-11}$ at 512 steps.
Lab-time RK4 then stalls at a roundoff floor between $2\times10^{-13}$ and $1\times10^{-12}$,
because the phase $\eta=t-z$ entering the field is formed by subtracting two
numbers of order $10^{3}$; the light-front scheme, whose orbit is algebraic and
whose independent variable is $\eta$ itself, continues to $2\times10^{-14}$ at
8192 steps. The Boris--BMT pusher converges at second order, from $0.91$ at
32 steps to $7\times10^{-6}$ at 8192 steps. The Higuera--Cary pusher, which
removes the spurious force that the Boris rotation produces for a relativistic
particle whose electric and magnetic forces nearly
cancel~\cite{Vay2008,HigueraCary2017}, also converges at second order but with
an error constant smaller by a factor of about $7\times10^{3}$, from $7\times10^{-5}$ to
$1\times10^{-9}$. Along this orbit $\ud\eta/\ud t=\kappa/\gamma$ varies only between
$4.3\times10^{-3}$ and $5.0\times10^{-3}$, so equal steps in $t$ and in $\eta$ resolve the carrier
equally well: the advantage of the light-front variables here is precision and
cost, not the removal of a stiffness.

\section{Numerical protocol for the intra-pulse polarization angle}
\label{sec:s5}

\subsection{Exact vanishing of the net rotation for linear polarization}

The rapidity formulation makes an exact statement possible before any
numerics. For a linearly polarized plane-wave pulse of compact support both
terms of main text Eq.~(4) are total derivatives, Eq.~(\ref{eq:sigmafull}), so
that
\begin{equation}
\Delta\Sigma\big|_{\eta\ge2\pi N}
=\Big[2\arctan\frac{a_{x}}{2}+a_{e}a_{x}\Big]_{0}^{2\pi N}=0
\qquad\text{for all } \varphi_{0},
\label{eq:invariantS}
\end{equation}
i.e., the electron leaves the pulse with exactly the polarization it entered
with, independently of the CEP, of $g$, of $a_{0}$ and of the envelope
shape~\cite{Ternov1968,Bagrov2014,Walser2002}. For elliptical polarization the
statement is replaced by the holonomy of Sec.~\ref{sec:s1ellip}. We verify
Eq.~(\ref{eq:invariantS}) numerically without imposing it: scanning the CEP
over 16 uniformly spaced values in $[0,2\pi)$ for an electron initially at rest
with spin along the propagation axis, $S^{\mu}(0)=(0,0,0,1)$, and $a_{0}=0.42$,
we find a residual net rotation of at most $3.2\times10^{-13}\deg$ for $N=2$ cycles and
$8.8\times10^{-13}\deg$ for $N=8$ cycles, consistent with the tolerance of the DOP853
reference integrator (relative tolerance $10^{-11}$, absolute tolerance
$10^{-13}$). An independent cross-check integrating the complete covariant
system (orbital four-velocity and spin together) in proper time $\tau$ with
DOP853 at the same tolerances gives a final rotation of at most $5.2\times10^{-13}\deg$ and
$|\gamma-1|\le2\times10^{-14}$ at the end of the pulse for $N=2$. This constraint has
an immediate physical consequence: any reported CEP-driven \emph{net}
polarization change for a plane-wave pulse must originate from an ingredient
outside the ideal plane-wave BMT description --- finite focusing, radiative
corrections, or a scattering event occurring while the pulse is still on.

\subsection{Pulse parameters used for the intra-pulse polarization curves}

The intra-pulse polarization curves presented in the main text are generated
using the $\cos^{2}$-envelope pulsed field
$a_{x}(\eta)=a_{0}\,\mathrm{Env}(\eta)\cos(\eta+\varphi_{0})$,
with the carrier frequency normalized to $\omega=1$ (natural units,
$c=m_{e}=|e|=1$) and the field amplitude fixed at $a_{0}=0.42$ throughout. The
envelope is $\mathrm{Env}(\eta)=\cos^{2}\!\big(\pi(\eta-T/2)/T\big)$ for
$\eta\in[0,T]$ with $T=2\pi N$, and zero outside this window; this
compactly supported form has continuous derivatives at the pulse edges
(avoiding spurious high-frequency content) and a single free duration
parameter $N$ (number of optical cycles), matching the pulse of the
convergence study of Sec.~\ref{sec:s3}. The electron is initialized at
rest, $u^{\mu}(0)=(1,0,0,0)$ ($\gamma_{0}=1$, light-front constant
$\kappa=\gamma_{0}-u_{z}(0)=1$), with spin along the propagation axis,
$S^{\mu}(0)=(0,0,0,1)$. For this initial condition the reference solution
coincides with the closed form of Eq.~(\ref{eq:closedS}) to $8\times10^{-13}$
at all phases.

Two quantities are extracted from each trajectory. The first is the
rest-frame polarization angle $\Sigma(\eta)$, obtained from the reference
solution through the pure boost to the instantaneous rest frame, and its peak,
\begin{equation}
\bm{\zeta}=\bm{S}-\frac{S^{0}}{\gamma+1}\,\bm{u}=(-\sin\Sigma,\,0,\,\cos\Sigma),
\qquad \Sigma^{\max}=\max_{\eta}|\Sigma(\eta)|;
\label{eq:thetaS}
\end{equation}
the angle between the spatial part of the laboratory four-vector $S^{\mu}$ and
$\hat z$ is frame dependent, is not a polarization, and is not used. The second,
the peak-energy spread, is the spread over CEP of the peak kinetic energy
$\gamma_{\max}-1$ relative to its CEP average,
\begin{equation}
\Delta \equiv \frac{\max_{\varphi_{0}}(\gamma_{\max}-1) - \min_{\varphi_{0}}(\gamma_{\max}-1)}{\langle\gamma_{\max}-1\rangle_{\varphi_{0}}}.
\label{eq:deltaKE}
\end{equation}
Both quantities are evaluated on a CEP grid of 49 uniformly spaced values in
$[0,2\pi]$ (the end points being equivalent) for each of $N=2$ and $N=8$ cycles.

\subsection{Extraction of the CEP-dependent spread}

For each $(\varphi_{0},N)$ pair, the BMT equation [main text Eq.~(3)] is
integrated in $\eta$ along the Volkov orbit, Eq.~(\ref{eq:volkov}), with an
eighth-order Dormand--Prince (DOP853) integrator at relative and absolute
tolerances $10^{-11}$ and $10^{-13}$, the same reference integrator used for
the exact-invariant check above and, at tighter tolerances, as the reference
of the convergence study of Sec.~\ref{sec:s3}. The orbital four-velocity is obtained algebraically
from the two exact plane-wave invariants ($k\cdot u=\text{const}$,
$u_{\perp}(\eta)=u_{\perp0}+a_{\perp}(\eta)-a_{\perp0}$), so only the BMT
spin equation is integrated numerically; this ensures the extraction of
$\Sigma(\eta)$ is limited solely by the spin-sector integration error,
not by any error in the orbital sector. The peak angle $\Sigma^{\max}$ is
recorded for each CEP value, and the spread
$\Delta\Sigma^{\max}=\max_{\varphi_{0}}\Sigma^{\max}-\min_{\varphi_{0}}\Sigma^{\max}$
is computed over the full CEP scan. Table~\ref{tab:excursion} summarizes the
results (reproducing the numbers plotted in Fig.~1(b)--(c) of the main
text): the spread of $\Sigma^{\max}$ falls from $3.02^{\circ}$ at $N=2$ to
$0.22^{\circ}$ at $N=8$ (a suppression factor of $13.8$), and the
peak-kinetic-energy spread falls from $26.6\%$ at $N=2$ to $1.90\%$ at
$N=8$; the latter ratio is independent of $a_{0}$ identically, since
$\gamma_{\max}-1=a_{\max}^{2}/2$ with $a_{\max}=\max_{\eta}|a_{x}|\propto a_{0}$.
Both spreads are the CEP dependence of $a_{\max}$, which ranges over
$0.365$--$0.420$ for $N=2$ and $0.416$--$0.420$ for $N=8$. The same scan
at $g=2$ gives spreads of $3.021^{\circ}$ and $0.219^{\circ}$, so the anomalous
moment contributes $0.004^{\circ}$ to the $N=2$ spread, and $\Sigma^{\max}$ coincides with
$2\arctan(a_{\max}/2)+a_{e}a_{\max}$ to $4\times10^{-12}\,{}^{\circ}$ on the grid.

\begin{table*}[!ht]
\caption{CEP-controlled peak rest-frame polarization angle and peak-energy spread for
$a_{0}=0.42$, electron initially at rest, scanned over 49 CEP values in
$[0,2\pi]$.}
\label{tab:excursion}
\begin{ruledtabular}
\begin{tabular}{lccc}
Pulse length & $\max_{\varphi_{0}}\Sigma^{\max}$ & $\min_{\varphi_{0}}\Sigma^{\max}$ & Spread $\Delta\Sigma^{\max}$ \\
\hline
$N=2$ & $23.75^{\circ}$ & $20.72^{\circ}$ & $3.02^{\circ}$ \\
$N=8$ & $23.75^{\circ}$ & $23.53^{\circ}$ & $0.22^{\circ}$ \\
\end{tabular}
\begin{tabular}{lcc}
Pulse length & $\gamma_{\max}$ range over CEP & $(\gamma_{\max}-1)$ spread $\Delta$ \\
\hline
$N=2$ & $1.0667$--$1.0882$ & $26.6\%$ \\
$N=8$ & $1.0865$--$1.0882$ & $1.90\%$ \\
\end{tabular}
\end{ruledtabular}
\end{table*}

\subsection{Coupling of the transient to a probe}

Because the net rotation vanishes identically [Eq.~(\ref{eq:invariantS})], a
CEP signature must be read while the pulse is still on: a probe acting at phase
$\eta_{s}$ sees the rest-frame polarization rotated about $\hat{\bm y}$ by
$\Sigma(\eta_{s})$, i.e.\ a spin-flip fraction $\sin^{2}[\Sigma(\eta_{s})/2]$
relative to the initial axis, whereas a probe acting after the pulse sees none.
We do not compute scattering cross sections; CEP-resolved nonlinear Compton
scattering~\cite{Mackenroth2010,Krajewska2012} is one route to such a
measurement.

\subsection{Statistical averaging and experimental comparability}

The curves above assume perfect control of the CEP and of the interaction phase
$\eta_{s}$. A comparison with a polarimetry measurement would require averaging
$\Sigma(\eta_{s};\varphi_{0})$ over the residual CEP jitter $\delta\varphi_{0}$
of the laser, for instance with a Gaussian kernel in $\varphi_{0}$, and over the
phase window sampled by the probe. No jitter-averaged numbers are reported here.

\section{Physics-informed neural network training details}
\label{sec:s6}

\subsection{Network architecture}

The network $N_{\Theta}(\eta)=(\gamma,u_{x},u_{y},u_{z},S^{0},S^{1},S^{2},S^{3})$
takes as input a Fourier-feature embedding of the light-front phase,
\begin{equation}
\mathbf{f}(\eta)=\big[\,\eta/T,\ \sin(k\omega\eta),\ \cos(k\omega\eta)\,\big]_{k=1}^{8}
\in\mathbb{R}^{17},
\label{eq:fourierfeat}
\end{equation}
which is passed to a multilayer perceptron with 5 hidden layers of 128 units
each and $\tanh$ activations throughout, implemented in \texttt{float64}
precision (PyTorch, CPU), for a total of $6.9\times10^{4}$ trainable
weights. The sinusoidal embedding counters the spectral bias of neural
networks~\cite{Rahaman2019,Tancik2020}, which in physics-informed networks makes
the low-frequency part of the solution converge first and slows the resolution
of several carrier cycles~\cite{Wang2021,Krishnapriyan2021}. The same encoding
was used in Ref.~\cite{AkintsovPPCF2026} for oscillatory spin-independent
dynamics; here its contribution is isolated by the controlled ablation of
Sec.~\ref{sec:s6acc}. Concatenating harmonics up to
$k_{\max}=8$---comfortably above the two fundamental cycles plus the
relativistic and BMT harmonic content actually present at $a_{0}=0.42$---eases
this bottleneck. We quantify the effect in
Sec.~\ref{sec:s6acc}. The output layer has 8 units, one for each of the Lorentz
factor $\gamma$, the three spatial four-velocity components
$(u_{x},u_{y},u_{z})$, and the four spin four-vector components
$(S^{0},S^{1},S^{2},S^{3})$. The $\tanh$ activation is used throughout
because the residual loss requires first derivatives with respect to $\eta$
obtained by automatic differentiation, and $\tanh$ is $C^{\infty}$ and
non-degenerate everywhere, unlike piecewise-linear activations whose second
derivative vanishes almost everywhere.

Initial conditions are imposed by a hard-constrained ansatz rather than as a
soft penalty: each raw network output is multiplied by a smooth switch
$\tanh(\eta/w)$ with $w=0.35$, which vanishes at $\eta=0$, plus additive
offsets, so that $u^{\mu}(0)=(1,0,0,0)$ and
$S^{\mu}(0)=(0,0,0,1)$ are satisfied exactly by construction; this was
verified by evaluating the network output at $\eta=0$ and confirming
agreement with the initial condition to machine precision, not merely to
the level of a soft penalty. The Lorentz factor is further parameterized as
$\gamma = 1+\tanh(\eta/w)\cdot\mathrm{softplus}(\cdot)$ which enforces
$\gamma\ge1$.

\subsection{Loss function and collocation sampling}

The loss combines the mean-squared residual of the orbital block
(equations for $\gamma$, $u_{x}$, $u_{y}$, $u_{z}$), the mean-squared
residual of the spin block (BMT equations for $S^{0},\dots,S^{3}$), and invariant
penalties enforcing the mass-shell condition $u\cdot u=1$ and the spin
normalization $S\cdot S=-1$; no labeled trajectory data enter the loss at
any point, consistent with main text Eq.~(7). Collocation points are
resampled every optimization step rather than fixed once at the start of
training, using $150$ uniform points, $150$ points in the pulse-central
region, and $50$ points near the pulse edges, over the physical domain
$\eta\in[0,T]$ with $T=2\pi N$; this resampling prevents the network from
overfitting to a static, finite set of collocation points. The orbital- and
spin-residual loss weights are ramped from $(3,1)$ to $(8,3)$ over the
first $15\,000$ steps to counteract the harder-to-fit orbital channel (in
particular $u_{x}$); the ramp is longer and smoother than in our initial
small-network trials, chosen to keep the larger Fourier-feature network
stable; the invariant penalty enters with the fixed weight $20$.

\subsection{Optimizer settings}

Training proceeds in two sequential stages. The first stage uses the Adam
optimizer with a cosine learning-rate schedule from
$\mathrm{lr}_{0}=10^{-3}$ down to $10^{-6}$, run for 32\,000 steps, with
gradient-norm clipping at $1.0$ and best-loss checkpointing (the parameter
state at the step of lowest observed total loss is retained, which guards
against the late-training loss oscillations characteristic of this
non-convex, high-frequency loss landscape). The second stage applies an
L-BFGS polish (1200 fixed collocation points, up to 3000 iterations, strong
Wolfe line search) to further reduce the residual once the Adam stage has
converged. Random seeds are fixed and recorded: \texttt{torch.manual\_seed(0)}
and \texttt{numpy.random.seed(0)} are set at the start of training, so no
random seed is left unset.

The network is trained for the same physical configuration used for the
main-text figures, $a_{0}=0.42$, $N=2$ cycles, $\varphi_{0}=0.7$~rad, with
the electron initially at rest (as in Sec.~\ref{sec:s5}, distinct from the
surfing-electron initial condition used for the field in the convergence
study of Sec.~\ref{sec:convergence}). Total training wall time was 3853~s
for the Adam stage plus 1387~s for the L-BFGS polish, with 8 CPU threads.
Table~\ref{tab:loss}
reports the loss trace at selected steps.

\begin{table*}[!ht]
\caption{PINN loss trace for $a_{0}=0.42$, $N=2$, $\varphi_{0}=0.7$~rad. Rows labeled by step give the weighted total loss and its terms on the resampled collocation batch of that step; rows marked $^{a}$ are evaluated on 1200 uniform points with the final weights $(8,3,20)$.}
\label{tab:loss}
\begin{ruledtabular}
\begin{tabular}{rcccc}
Step & Total loss & Orbital MSE & Spin MSE & Invariant MSE \\
\hline
0 & $3.27\times10^{1}$ & $6.62\times10^{-2}$ & $2.06\times10^{-2}$ & $1.63\times10^{0}$ \\
2000 & $2.15\times10^{-4}$ & $3.77\times10^{-5}$ & $1.40\times10^{-5}$ & $2.95\times10^{-6}$ \\
8000 & $2.51\times10^{-5}$ & $1.18\times10^{-6}$ & $6.85\times10^{-7}$ & $8.49\times10^{-7}$ \\
16000 & $4.70\times10^{-5}$ & $1.72\times10^{-6}$ & $1.71\times10^{-6}$ & $1.41\times10^{-6}$ \\
24000 & $6.23\times10^{-6}$ & $1.93\times10^{-7}$ & $7.31\times10^{-8}$ & $2.23\times10^{-7}$ \\
31999 & $1.05\times10^{-6}$ & $8.73\times10^{-8}$ & $8.74\times10^{-9}$ & $1.61\times10^{-8}$ \\
Adam, best checkpoint$^{a}$ & $1.04\times10^{-6}$ & $8.59\times10^{-8}$ & $1.06\times10^{-8}$ & $1.59\times10^{-8}$ \\
after L-BFGS polish$^{a}$ & $1.02\times10^{-6}$ & $8.45\times10^{-8}$ & $9.87\times10^{-9}$ & $1.59\times10^{-8}$ \\
\end{tabular}
\end{ruledtabular}
\end{table*}

\subsection{Accuracy against the reference integrator}
\label{sec:s6acc}

The trained network is compared against the DOP853 reference on a
2000-point dense grid over $\eta\in[0,T]$. The maximum absolute error over
this grid is $4.73\times10^{-4}$ for $\gamma(\eta)$ and $3.90\times10^{-5}$
for the spin components $S^{\mu}(\eta)$ (maximum over $\mu$), with
individual-component errors $3.57\times10^{-5}$ ($S^{0}$),
$3.90\times10^{-5}$ ($S^{1}$), $1.44\times10^{-5}$ ($S^{2}$), and
$1.73\times10^{-5}$ ($S^{3}$). Against the closed form, Eq.~(\ref{eq:closedS}), the spin error is
$3.90\times10^{-5}$, since the reference and the closed form differ by less than $10^{-12}$. Figure~2(d) of the main text shows the
error traces $|\Delta\gamma(\eta)|$ and $\max_{\mu}|\Delta S^{\mu}(\eta)|$
across the pulse, with isolated deep minima where the network curve crosses the reference
and the error changes sign.

We emphasize which ingredient is responsible. In a controlled ablation the same network (5 hidden
layers of 128 units), loss, collocation sampling, Adam schedule and L-BFGS polish are used with the
Fourier-feature embedding removed ($k_{\max}=0$, input $\eta/T$ only). The ablated network reaches
$8.11\times10^{-4}$ in $\gamma$ and $3.94\times10^{-4}$ in the spin components, larger by factors of $1.7$ and $10$
than with the embedding: at two carrier cycles the spectral bias slows convergence, mostly in the spin
sector, but does not prevent it. A smaller plain-input network (4 hidden layers of 64 units, 24\,000 Adam steps, no L-BFGS)
reaches $5.59\times10^{-2}$ and $2.24\times10^{-3}$. The L-BFGS polish changes the maximum absolute errors of the
main network from $4.72\times10^{-4}$ to $4.73\times10^{-4}$ in $\gamma$ and from $4.90\times10^{-5}$ to
$3.90\times10^{-5}$ in the spin components. The residual gap to the $10^{-11}$ tolerance of the reference
integrator is the expected price of a mesh-free global representation.

\subsection{Elliptical polarization}
\label{sec:s6ell}

\begin{figure*}[!ht]
\centering
\includegraphics[width=0.55\textwidth]{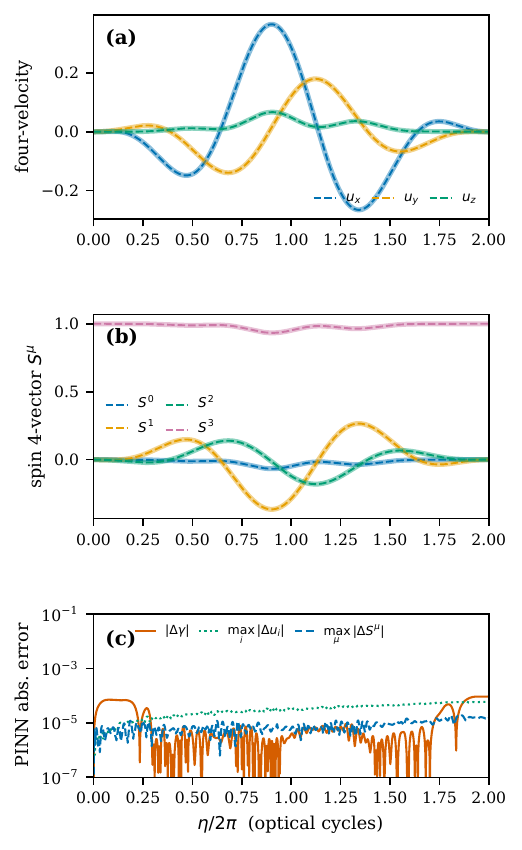}
\caption{PINN solution for an elliptically polarized pulse ($\delta=0.5$, $a_{0}=0.42$, $N=2$,
$\varphi_{0}=0.7$~rad; electron initially at rest, spin along $\hat z$), for which no closed-form solution is known.
(a) Spatial four-velocity components and (b) spin four-vector components from the PINN (thin dashed)
against the DOP853 reference (thick light). (c) Pointwise absolute errors of the network.}
\label{fig:s1}
\end{figure*}

To test the solver in a case for which no closed-form solution is known, we train the same network, with the same loss,
collocation sampling and optimizer schedule, for an elliptically polarized pulse
$a_{x}=a_{0}\,\mathrm{Env}(\eta)\cos(\eta+\varphi_{0})/\sqrt{1+\delta^{2}}$,
$a_{y}=\delta\,a_{0}\,\mathrm{Env}(\eta)\sin(\eta+\varphi_{0})/\sqrt{1+\delta^{2}}$ with $\delta=0.5$,
$a_{0}=0.42$, $N=2$ and $\varphi_{0}=0.7$~rad; the only change is the field entering the residuals.
Against the DOP853 reference on the 2000-point grid the network reaches maximum absolute errors of
$9.37\times10^{-5}$ in $\gamma$, $6.27\times10^{-5}$ in the spatial four-velocity and $1.89\times10^{-5}$ in the
spin components (Fig.~\ref{fig:s1}), with $|u\cdot u-1|\le2\times10^{-4}$ and $|S\cdot S+1|\le2\times10^{-5}$; training took
4601~s for the Adam stage and 512~s for the L-BFGS polish.

\subsection{Behavior of the invariant penalties}

The invariant penalties are not decorative. Their contribution to the total
loss is reported in the last column of Table~\ref{tab:loss}: it falls from
$1.625$ at initialization to $1.6\times10^{-8}$ at the best Adam
checkpoint and $1.6\times10^{-8}$ after the L-BFGS polish (both on 1200 uniform points), so that
the mass-shell condition $u\cdot u=1$ and the spin normalization $S\cdot S=-1$ are satisfied by
the trained network to $9\times10^{-4}$ and $3\times10^{-5}$, respectively, at every point of a 2000-point grid
across the whole pulse, without either invariant
ever being enforced algebraically. Because the residual and invariant terms
are optimized jointly, this also acts as an independent consistency check on
the solution: a network that fitted the residuals but violated the
invariants would be exhibiting the classic mode of PINN failure in which the
residual is small on the collocation set but the solution has drifted off the
physical manifold between collocation points.

We did not carry out a controlled ablation in which the invariant penalties
are removed and the network retrained under an otherwise identical budget;
such a study would quantify the individual contribution of each penalty and
is a natural extension, but no ablation numbers are reported here and none
should be inferred from the above.

\section{Reproducibility}
\label{sec:s7}

\subsection{Units and physical parameters}

All results in this Supplemental Material and in the main text use natural
units $c=m_{e}=|e|=1$, with electron charge-to-mass ratio $q/m=-1$. The
propagation direction is $+z$, and the light-front phase is
$\eta=\omega(t-z)$ with carrier frequency $\omega=1.0$ (normalized). The
vector potential is $a_{x}(\eta)=a_{0}\,\mathrm{Env}(\eta)\cos(\eta+\varphi_{0})$
with $a_{y}=0$ (linear polarization along $x$), and the envelope is the
$\cos^{2}$ form of Sec.~\ref{sec:s5}. The field amplitude used throughout
is $a_{0}=0.42$. The anomalous magnetic moment used in the BMT equation is
$a_{e}=0.00115965218$ (electron $(g-2)/2$), with $g/2=1+a_{e}$. The
reference (``ground truth'') integrator for all comparisons is
\texttt{scipy.integrate.solve\_ivp} with \texttt{method="DOP853"},
\texttt{rtol=1e-11}, \texttt{atol=1e-13} (tightened to $10^{-13}$ and
$10^{-15}$ for the reference of the convergence study of Sec.~\ref{sec:s3}), integrated in the light-front
variable $\eta$; the orbital four-velocity is obtained algebraically from
the two exact plane-wave invariants, and only the BMT spin equation is
integrated numerically.

\subsection{Software, code, and random seeds}

All numerical results were produced with Python 3.14.3,
\texttt{scipy 1.18.1}, \texttt{numpy 2.4.2}, \texttt{matplotlib 3.11.1}, and
\texttt{torch 2.10.0+cpu} (CPU build). The reference ODE integrations are fully deterministic
and involve no random seed. For the PINN training of Sec.~\ref{sec:s6},
\texttt{numpy.random.seed(0)} and \texttt{torch.manual\_seed(0)} are set at
the start of training and are the only sources of stochasticity in the
pipeline (governing collocation-point resampling and weight initialization).
Figures are produced as vector PDFs at APS single-column width (3.4 in)
using an Okabe--Ito colorblind-safe palette and an 8--9~pt serif font, with
panel labels and no embedded titles. The complete code, the computed data and
the trained networks are deposited as release 1.0.1 in the Zenodo repository of
Ref.~\cite{Akintsov2026code} (source at \url{https://github.com/NewArtY/lfspin}); its driver
\texttt{reproduce\_all.py} regenerates every number, table and figure of the Letter and of
this Supplemental Material and checks the deposited files against their SHA-256 manifest.

\end{document}